\documentclass[aps,prd,preprint,showkeys]{revtex4}

\usepackage{latexsym}
\usepackage{graphicx}

\begin{document}


\title{Surface gravity and Hawking temperature \\ from entropic force
  viewpoint}

\author{Ee Chang-Young}
\email[]{cylee@sejong.ac.kr}

\author{Myungseok Eune}
\email[]{younms@sejong.ac.kr}

\author{Kyoungtae Kimm}
\email[]{helloktk@naver.com}
\affiliation{Department of Physics and Institute of Fundamental
  Physics, \\ Sejong University, Seoul 143-747, Korea}

\author{Daeho Lee}
\email[]{dhleep@sogang.ac.kr}
\affiliation{Basic Science Research Institute, Sogang University,
	Seoul 121-742, Korea}



\begin{abstract}
  We consider a freely falling holographic screen for the
  Schwarzschild and Reissner-Nordstr\"om black holes and evaluate the
  entropic force \`a la Verlinde.  When the screen crosses
  the event horizon, the temperature of the screen agrees to the
  Hawking temperature and the entropic force gives rise to the surface
  gravity for both of the black holes.
\end{abstract}



\maketitle


Hawking radiation results from the quantum effect of fields in a
classical geometry with an event horizon \cite{hawking}. Thus the
study of Hawking radiation may provide a clue in constructing the
theory of quantum gravity.  Considering its importance, it may be
useful to have various interpretations of the effect from different
angles, which may lead to a further understanding of the nature of
black holes.

There have been several ways of providing the interpretation of
Hawking radiation. One of them is the proposal suggested by Robinson
and Wilczek \cite{rw}, which is that Hawking radiation plays the role
of preserving general covariance at the quantum level by canceling the
diffeomorphism anomaly at the event horizon(see also
\cite{Banerjee:2007uc,Banerjee:2008az}).  The flux of Hawking
radiation can be also obtained through the scattering analysis and
there have been the studies of the grey body factor for various black
holes to calculate the Hawking temperature
\cite{gl,gk:grey,ms:grey,km,cl,lm,ko:ads,koy,cfm,ko}.

Recently, it has been suggested by Verlinde that gravitational
interaction can be interpreted as a kind of entropic force through the
holographic principle and the equipartition rule~\cite{verlinde}. This
entropic formulation of inertia and gravity has been used to study
thermodynamics at the apparent horizon of the
Friedmann-Robertson-Walker universe~\cite{Shu:2010nv}, Friedmann
equations~\cite{Cai:2010hk}, Newtonian gravity in loop quantum
gravity~\cite{Smolin:2010kk}, holographic dark
energy~\cite{Li:2010cj,efs,danielson}, an extension to Coulomb
force~\cite{Wang:2010px}, entropic corrections to Newton's
law~\cite{Modesto:2010rm}, the Plank scale effect~\cite{Ghosh:2010hz},
and the gauge/gravity duality~\cite{Zhao:2010vt}. There have been many
works for the entropic force in the cosmological models
\cite{Gao:2010fw,zgz,wang:y,Wei:2010ww,Ling:2010zc,Easson:2010xf} and
the black hole backgrounds
\cite{Myung:2010jv,mk,Liu:2010na,Cai:2010sz,Tian:2010uy,Kuang:2010gs}. In
this paper, we study the Hawking temperature and surface gravity by
calculating the entropic force on the holographic screen for
spherically symmetric and static black holes.

\begin{figure}[pbt]
  \centering
  \includegraphics[width=0.4\textwidth]{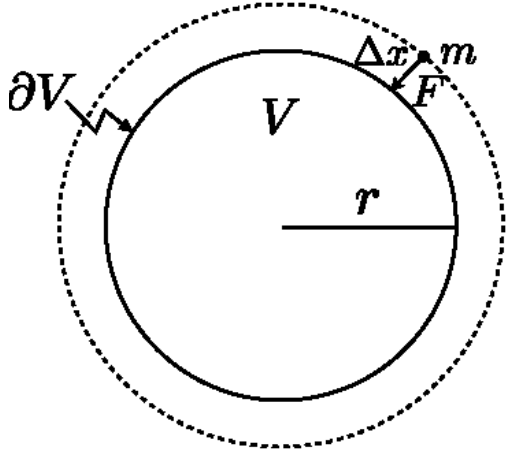}
  \caption{A test particle with mass $m$ approaches the holographic
    screen, which is the boundary $\partial V$ of the sphere with
    volume $V$ with radius $r$. $\Delta x$ is the distance that the
    particle moves.}
  \label{fig:system}
\end{figure}

We consider a test particle with mass $m$ which approaches
the holographic screen from a distance $\Delta x$ (see
FIG.~\ref{fig:system}). In this case the change of entropy associated
with the information on the boundary is assumed to be~\cite{verlinde}
\begin{equation}
  \label{dS}
  \Delta S = 2\pi m \Delta x,
\end{equation}
where we have taken the natural units of $c = \hbar = k_B = 1$. From
the thermodynamic laws, the entropic force $F$ acting on the test
particle is given by~\cite{verlinde,Hossenfelder:2010ih}
\begin{equation}
  \label{F:def}
  F \Delta x = T \Delta S,
\end{equation}
where $T$ is the temperature on the screen.  Substituting
Eq.~(\ref{dS}) into Eq.~(\ref{F:def}), the entropic force can be
written as
\begin{equation}
  \label{F:T}
  F = 2\pi m T.
\end{equation}

It is assumed that the total number $N$ of the fundamental bits is
proportional to the area $A$ of the screen, \textit{i.e.,}
\begin{equation}
  \label{N:def}
  N = \frac{A}{G},
\end{equation}
where $G$ is a given constant for the moment and will be identified as
the Newton's constant later on. When a total energy $E$ is divided
uniformly over the bits $N$, the temperature is determined by the
equipartition rule \cite{verlinde}
\begin{equation}
  \label{equipartition}
  E = \frac12 NT,
\end{equation}
as the average energy per bit.

Now, we consider the Schwarzschild-like black holes which include both
Schwarzschild and Reissner-Nordstr\"om (RN) black holes with the line
element given by
\begin{equation}
  \label{metric:SS}
  ds^2 = g_{\mu\nu} dx^\mu dx^\nu = - f(r) dt^2 + \frac{dr^2}{f(r)} + r^2 d\Omega_2^2,
\end{equation}
where $f(r)$ vanishes at the event horizon $r_H$.

In order to consider a freely falling holographic screen, we introduce
the Painlev\'e-Gullstrand (PG) line element~\cite{visser} of the
Schwarzschild-like geometry (\ref{metric:SS}). The ingoing ($+$) and
outgoing ($-$) PG coordinates can be written as
\begin{equation}
  \label{PG:1st}
  ds^2 = - d\tau^2 + \left( dr \pm \sqrt{1-f} d\tau \right)^2 +
  r^2 d\Omega_2^2,
\end{equation}
or, equivalently,
\begin{equation}
  \label{PG:2nd}
  ds^2 = - f\, d\tau^2 \pm
  2\sqrt{1-f} d\tau dr + dr^2 + r^2 d\Omega_2^2.
\end{equation}
Note that constant-time slices are completely flat in this coordinates
system. The PG coordinates are related to the Schwarzschild-like
coordinates by
\begin{equation}
  \label{tau:t:2nd}
  d\tau = dt \pm \frac{\sqrt{1-f}}{f} dr.
\end{equation}

The conserved energy of the gravitational field over a spacelike
hypersurface $V$ at a certain time is given by the Komar
integral~\cite{carroll}
\begin{equation}
  \label{E:def}
  E(V) = \frac{1}{4\pi G} \oint_{\partial V} \nabla^\mu \xi^\nu n_\mu
  \sigma_\nu dA,
\end{equation}
where $n_\mu$ and $\sigma_\mu$ are the unit normal vectors
perpendicular to the hypersurface $V$ and to its boundary $\partial V$
with constant radius $r$, respectively. The normal vectors satisfy
$n_\mu n^\mu = -1$ and $\sigma_\mu \sigma^\mu = 1$. And $\xi^\alpha$
is a timelike Killing vector field which satisfies the Killing
equation of $\nabla_\mu \xi_\nu + \nabla_\nu\xi_\mu = 0$.

For $r>r_H$, the Killing vector and the unit normal vectors are given
by $\xi^\mu = \delta^\mu_\tau$, $n_\mu = -\delta^\tau_\mu$, and
$\sigma^\mu = \delta^\mu_r$ in the PG coordinates.  Then, from
Eq.~(\ref{E:def}), the energy is given by
 \begin{equation}
   \label{E}
   E = \frac{r^2}{2G} f'(r),
 \end{equation}
 where ${}'$ denotes a differentiation with respect to $r$.  The entropy
 defined on the freely falling holographic screen located outside the
 horizon is given by
\begin{equation}
  \label{S:def}
  S = \frac{A}{4G}.
\end{equation}
The above relation can be restated as
\begin{equation}
  \label{N:S}
  N = 4S
\end{equation}
via Eq.~(\ref{N:def}).  Using the equipartition rule for the energy
$E$ in Eq.~(\ref{equipartition}) and the relation (\ref{N:S}), we get
the following relation~\cite{Padmanabhan:2003pk,Padmanabhan:2009vy}
\begin{equation}
  \label{equipartition:local}
  T = \frac{E}{2S}.
\end{equation}
Then, from Eqs.~(\ref{E}) and (\ref{S:def}) the temperature on the
holographic screen with the constant radius $r$ is given by
\begin{equation}
  \label{T}
  T(r) = \frac{1}{4\pi} f'(r).
\end{equation}
At the horizon, the surface gravity $\kappa$ is defined with the Killing vector
$\xi^\mu$ by the following relation
\begin{equation}
   \xi^\mu \nabla_\mu \xi^\alpha =
   \kappa \xi^\alpha. \label{kappa:def}
\end{equation}
In the Schwarzschild-like geometry, $\kappa=\frac12 f'(r_H)$.  Thus,
the temperature (\ref{T}) evaluated at the horizon ($r=r_H$) is
related to the surface gravity as follows:
\begin{equation}
  \label{T:H}
  T(r_H) = \frac{1}{4\pi} f'(r_H) = \frac{\kappa}{2\pi}.
\end{equation}
Since the Hawking temperature $T_H$ is related to the surface gravity
by $T_H = \frac{\kappa}{2\pi}$ \cite{gh}, the temperature at
the horizon $T(r_H)$ is the same as the Hawking temperature.

Finally, the entropic force is obtained from Eqs.~(\ref{F:T}) and
(\ref{T}) as
\begin{equation}
  \label{F:result}
  F = 2\pi m T(r) = \frac12 m f'(r),
\end{equation}
where $m$ is the mass of the test particle. For the Schwarzschild case
$f(r) = 1 - 2GM/r$, the entropic force becomes the Newton force
\begin{equation}
  \label{F:Sch}
  F = \frac{GmM}{r^2} ,
\end{equation}
where $M$ is the mass of the black hole and $G$ is identified as the
Newton's gravitational constant. For the RN case $f(r) = 1 -
2GM/r + GQ^2/r^2$, the entropic force becomes
 \begin{equation}
   \label{F:RN}
   F= \frac{Gm M}{r^2} - \frac{Gm Q^2}{r^3},
\end{equation}
where $M$ and $Q$ are the mass and charge of the black hole,
respectively. Note that the force in Eq.~(\ref{F:RN}) contains
the gravitational effect only, since the test particle's charge
does not play any role in this consideration. Via the metric, only the geometry
which is determined by the energy of the system
plays a role here.
At the event horizon($r=r_+$), the entropic
force in the RN black hole becomes the usual expression
\begin{displaymath}
  F=\frac{m(r_+ - r_-)}{2r_+^2} =  m\kappa,
\end{displaymath}
where $r_\pm \equiv GM \pm \sqrt{G^2 M^2 - G Q^2}$.

In Ref.~\cite{Wang:2010px}, the RN black hole was considered. There
the motivation came from a mismatch between the Hawking temperature
and the energy in the entropic force formulation. However, the energy
used there contains only the gravitational mass not including the
electromagnetic energy. On the other hand, we obtain the energy
using the Komar integral with the RN metric, which contains both mass
and charge contributions to the energy. This yields the correct
relation in our case.

\begin{acknowledgments}
  We thank Youngone Lee for helpful discussions. This work was
  supported by the National Research Foundation (NRF) of Korea grants
  funded by the Korean government (MEST) [R01-2008-000-21026-0 and
  NRF-2009-0075129 (E.\ C.-Y.\ and K.\ K.), NRF-2009-351-C00109 (M.\
  E.), and NRF-2009-351-C00111 (D.\ L.)].
\end{acknowledgments}



\begin{thebibliography}{99}

\bibitem{hawking}
  S.~W.~Hawking,
  Commun.\ Math.\ Phys.\  {\bf 43} (1975) 199
  [Erratum-ibid.\  {\bf 46} (1976) 206].


\bibitem{rw}
  S.~P.~Robinson and F.~Wilczek,
  Phys.\ Rev.\ Lett.\  {\bf 95} (2005) 011303
  [arXiv:gr-qc/0502074].



\bibitem{Banerjee:2007uc}
  R.~Banerjee and S.~Kulkarni,
  Phys.\ Lett.\  B {\bf 659} (2008) 827
  [arXiv:0709.3916 [hep-th]];
  Phys.\ Rev.\  D {\bf 79} (2009) 084035
  [arXiv:0810.5683 [hep-th]].

\bibitem{Banerjee:2008az}
  R.~Banerjee,
  Int.\ J.\ Mod.\ Phys.\  D {\bf 17} (2009) 2539
  [arXiv:0807.4637 [hep-th]].




\bibitem{gl}
  K.~Ghoroku and A.~L.~Larsen,
  Phys.\ Lett.\  B {\bf 328} (1994) 28
  [arXiv:hep-th/9403008].

\bibitem{gk:grey}
  S.~S.~Gubser and I.~R.~Klebanov,
  Phys.\ Rev.\ Lett.\  {\bf 77} (1996) 4491
  [arXiv:hep-th/9609076].

\bibitem{ms:grey}
  J.~M.~Maldacena and A.~Strominger,
  Phys.\ Rev.\  D {\bf 55} (1997) 861
  [arXiv:hep-th/9609026].

\bibitem{km}
  I.~R.~Klebanov and S.~D.~Mathur,
  Nucl.\ Phys.\  B {\bf 500} (1997) 115
  [arXiv:hep-th/9701187].

\bibitem{cl}
  M.~Cvetic and F.~Larsen,
  Nucl.\ Phys.\  B {\bf 506} (1997) 107
  [arXiv:hep-th/9706071].

\bibitem{lm}
  H.~W.~Lee and Y.~S.~Myung,
  Phys.\ Rev.\  D {\bf 58} (1998) 104013
  [arXiv:hep-th/9804095].

\bibitem{ko:ads}
  W.~T.~Kim and J.~J.~Oh,
  Phys.\ Lett.\  B {\bf 461} (1999) 189
  [arXiv:hep-th/9905007].

\bibitem{koy}
  W.~T.~Kim, J.~J.~Oh and K.~H.~Yee,
  Phys.\ Rev.\  D {\bf 66} (2002) 044017
  [arXiv:hep-th/0201117].

\bibitem{cfm}
  G.~Clement, J.~C.~Fabris and G.~T.~Marques,
  Phys.\ Lett.\  B {\bf 651} (2007) 54
  [arXiv:0704.0399 [gr-qc]].

\bibitem{ko}
  W.~Kim and J.~J.~Oh,
  J.\ Korean Phys.\ Soc.\  {\bf 52} (2008) 986
  [arXiv:0709.1754 [hep-th]].



\bibitem{verlinde}
  E.~P.~Verlinde,
  arXiv:1001.0785 [hep-th].



\bibitem{Shu:2010nv}
  F.~W.~Shu and Y.~Gong,
  arXiv:1001.3237 [gr-qc].

\bibitem{Cai:2010hk}
  R.~G.~Cai, L.~M.~Cao and N.~Ohta,
  arXiv:1001.3470 [hep-th].

\bibitem{Smolin:2010kk}
  L.~Smolin,
  arXiv:1001.3668 [gr-qc].

\bibitem{Li:2010cj}
  M.~Li and Y.~Wang,
  arXiv:1001.4466 [hep-th].

\bibitem{efs}
D.~A.~Easson, P.~H.~Frampton and G.~F.~Smoot,
  arXiv:1002.4278 [hep-th].

\bibitem{danielson}
 U.~H.~Danielsson,
  arXiv:1003.0668 [hep-th].

\bibitem{Wang:2010px}
  T.~Wang,
  arXiv:1001.4965 [hep-th].

\bibitem{Modesto:2010rm}
  L.~Modesto and A.~Randono,
  arXiv:1003.1998 [hep-th].

\bibitem{Ghosh:2010hz}
  S.~Ghosh,
  arXiv:1003.0285 [hep-th].

\bibitem{Zhao:2010vt}
  Y.~Zhao,
  arXiv:1002.4039 [hep-th].






\bibitem{Gao:2010fw}
  C.~Gao,
  arXiv:1001.4585 [hep-th].

\bibitem{zgz}
 Y.~Zhang, Y.~g.~Gong and Z.~H.~Zhu,
  arXiv:1001.4677 [hep-th].

\bibitem{wang:y}
 Y.~Wang,
  arXiv:1001.4786 [hep-th].

\bibitem{Wei:2010ww}
  S.~W.~Wei, Y.~X.~Liu and Y.~Q.~Wang,
  arXiv:1001.5238 [hep-th].

\bibitem{Ling:2010zc}
  Y.~Ling and J.~P.~Wu,
  arXiv:1001.5324 [hep-th].

\bibitem{Easson:2010xf}
  D.~A.~Easson, P.~H.~Frampton and G.~F.~Smoot,
  arXiv:1003.1528 [hep-th].

\bibitem{Myung:2010jv}
  Y.~S.~Myung,
  arXiv:1002.0871 [hep-th].

\bibitem{mk}
 Y.~S.~Myung and Y.~W.~Kim,
  arXiv:1002.2292 [hep-th].

\bibitem{Liu:2010na}
  Y.~X.~Liu, Y.~Q.~Wang and S.~W.~Wei,
  arXiv:1002.1062 [hep-th].

\bibitem{Cai:2010sz}
  R.~G.~Cai, L.~M.~Cao and N.~Ohta,
  arXiv:1002.1136 [hep-th].

\bibitem{Tian:2010uy}
  Y.~Tian and X.~Wu,
  arXiv:1002.1275 [hep-th].

\bibitem{Kuang:2010gs}
  X.~Kuang, Y.~Ling and H.~Zhang,
  arXiv:1003.0195 [gr-qc].


\bibitem{Hossenfelder:2010ih}
  S.~Hossenfelder,
  arXiv:1003.1015 [gr-qc].



\bibitem{visser}
    M.~Visser,
  Class.\ Quant.\ Grav.\  {\bf 15} (1998) 1767
  [arXiv:gr-qc/9712010].

\bibitem{carroll}
  S.~M.~Carroll,
 {\it Spacetime and geometry: An introduction to general relativity}
   (Addison-Wesley, San Francisco, 2004).

\bibitem{Padmanabhan:2003pk}
  T.~Padmanabhan,
  Class.\ Quant.\ Grav.\  {\bf 21} (2004) 4485
  [arXiv:gr-qc/0308070].

\bibitem{Padmanabhan:2009vy}
  T.~Padmanabhan,
  arXiv:0911.5004 [gr-qc];
  arXiv:0912.3165 [gr-qc].

\bibitem{gh}
  G.~W.~Gibbons and S.~W.~Hawking,
  Phys.\ Rev.\  D {\bf 15} (1977) 2752.

\bibitem{Caravelli:2010be}
  F.~Caravelli and L.~Modesto,
  arXiv:1001.4364 [gr-qc].

\end{thebibliography}

\end{document}